\documentclass{elsart}

\usepackage{lscape}
\usepackage{epsfig}
\usepackage{amssymb}
\journal{NIMB}

\begin{document}
\begin{frontmatter}

\title{Verification of the content, isotopic composition and age of plutonium in Pu-Be neutron sources by gamma-spectrometry}

\author{Cong Tam Nguyen}
\ead{tam@iki.kfki.hu}
\address {{\small \it Institute of Isotopes of the Hungarian Academy of Sciences}\\
{\small \it H-1525 Budapest, P.O. Box 77, Hungary} }

\begin{abstract}
A non-destructive, gamma-spectrometric method for verifying the plutonium content of Pu-Be neutron sources has been developed. It is also shown that the isotopic composition and the age of plutonium (Pu) can be determined in the intensive neutron field of these sources by the ``Multi-Group Analysis'' method. Gamma spectra were taken in the far-field of the sample, which was assumed to be cylindrical. The isotopic composition and the age of Pu were determined using a commercial implementation of the Multi-Group Analysis algorithm. The Pu content of the sources was evaluated from the count rates of the gamma-peaks of $^{239}$Pu, relying on the assumption that the gamma-rays are coming to the detector parallel to each other. The determination of the specific neutron yields and the problem of neutron damage to the detector are also discussed.
\end{abstract}

\end{frontmatter}

\section{Introduction}

In recent years Pu-Be sources have been intensively studied \cite{CBM-ESARDA,CBM-NIM,CBM-OAH} because they represent a nuclear safety and safeguards issue. A large number of such sources (around 200) - mostly out of use - are stored in Hungary (and this is the case in several neighbouring countries as well). The Pu-content of these sources is to be accounted for, regularly reported to and inspected by the International Atomic Energy Agency, and from 2004 on, they are also subject to EURATOM safeguards. Because the late Soviet supplier did not declare the Pu-content of the sources, the values for the Pu content presently on record are based on rough overestimates assuming that the sources contain pure $^{239}$Pu.

Using the Multi-Group Analysis algorithm \cite{MGA,MGA++} to analyze the gamma spectra of some Pu-Be sources, however, has shown that the $^{239}$Pu component varies in the range of 75-95\%. It is known that the specific neutron yields can be very different for different Pu isotopes, for example, $\sim10^4$ n/sg for $^{239}$Pu, and $\sim10^7$ n/sg for $^{238}$Pu as calculated \cite{CBM-NIM} from specific alpha yield \cite{IAEA-safety,PANDA}. Consequently, the total Pu content should be corrected by using the following data: 1. the measured Pu isotopic composition, 2. neutron output from declared or measured data and 3. specific yield values. The calculating procedure which uses these corrections we call  ``combined method'' or ``CBM'' and its results were reported in Refs. \cite{CBM-ESARDA,CBM-NIM,CBM-OAH}.
In case of lower $^{239}$Pu isotope ratio (about 70-80\%) the total Pu contents are substantially lower than the values still kept in the files for safeguards. A typical example is a source with 24 g of Pu found by the CBM instead of the value of 178 g on file. However, since specific neutron yields are not isotopic physical parameters, they are strongly dependent on the technology applied by the manufacturer. For example, the value for the specific neutron yield of $^{239}$Pu may vary in the range $(1.2-6.2)\times 10^4$ n/sg, \cite{PANDA}. Therefore, the combined method may contain systematical errors because of the lack of the precise knowledge of the Pu specific neutron yields. The Pu contents estimated by the CBM should be considered as upper limits for the Pu content \cite{CBM-NIM}. Similarly, the lower limits can be estimated if the maximum specific neutron yields are used.  Although these estimates are subject to the high uncertainty associated with the specific neutron yields needed for the calculation of the Pu-content, the
CBM gave an indication that the Pu contents of Pu-Be neutron sources must be carefully investigated.  

In parallel with the gamma-spectrometric measurement of the isotopic composition, a high-resolution gamma-spectrometric method which does not require the knowledge of the specific neutron yields was developed for the experimental determination of the plutonium content of the Pu-Be sources. Furthermore, combining the results of the gamma-spectrometric measurement with the measurement of the total neutron output the specific neutron yields of the sources can also be determined, thus providing additional information about the sources. 

During developing this topic, another approach using neutron coincidence counting technique was considered for the estimation of the Pu content.  A neutron coincidence counter was built and a correlation was obtained between the Pu content and the ratio of the number of real coincidences ($R$) to the total number of counts ($T$). This method will be called here the ``R/T'' method. An important advantage of this method is that there is no need for using the Pu isotopic composition measured by gamma spectrometry. However, to develop the R/T method standard Pu-Be neutron sources of known isotopic composition and Pu content are needed for getting an accurate calibration curve. Unfortunately, there are no such sources in our laboratory and, as far as we know, there are none in the laboratories with which we made contact. The gamma-spectrometric method described here made it possible to prepare a set of calibrating sources which can be used for developing the R/T method. The results of this research can help to develop the R/T method and to monitor the specific neutron yield of the sources. 

This paper presents a procedure for evaluating the Pu contents of Pu-Be neutron sources based purely on high-resolution gamma-spectrometry.  Pu-Be sources with neutron output from $10^4$ n/s to $10^7$ n/s were assayed. The samples were set up in the far field of the detector, so that the gamma-rays may be assumed to be  coming to the detector parallel to its axis. As the Pu-Be sources are usually encapsulated in steel cylinders of known wall thickness, the correction factor for self-absorption can be appropriately calculated by a numerical model for the far-field assay of a cylindrical sample. 

Note that the Pu content can also be derived from the count rates of the gamma-rays of $^{239}$Pu by a method called the ``infinite energy method'' \cite{infinite}. For Pu-Be sources, however, the uncertainty of that method may reach up to 20-25 \%. If, however, the self-absorption correction can be accuretely calcualted, as is the case with the samples assayed in this work, the precision of determining the Pu content by the procedure described in this paper is expected to be better than the precision of the infinite energy method.

\section{Determination of the isotopic composition and of the age of plutonium}

\begin{figure}[htbp]
\begin{center}
\epsfig{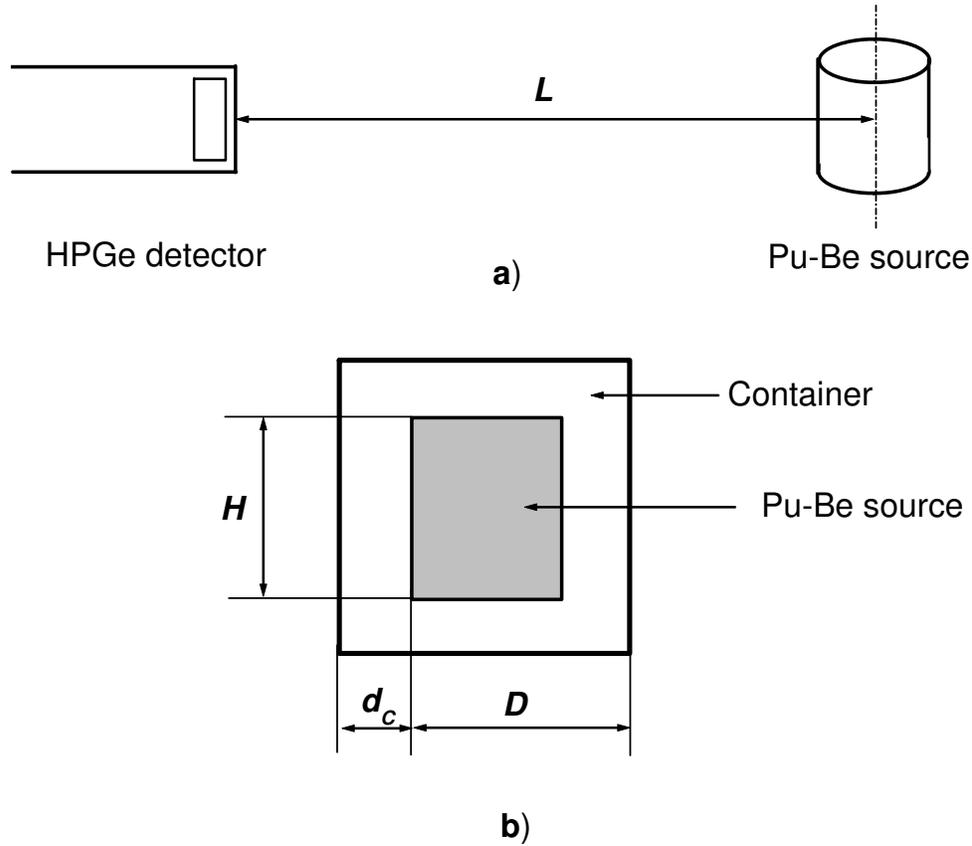}
\caption{The measurement setup}
\label{setup}
\end{center}
\end{figure}

Spectra of Pu-Be neutron sources were taken in the 0-600 keV energy region by a 2000 mm$^2$ planar high-purity germanium detector (Canberra GL2020R) at 20-200 cm source-to-detector distance (see Fig. \ref{setup}). The parameters of the sources and of the measurements are given in Table \ref{Tab:parameters}. The multi-channel analyzer was set up at 8196 channels to get the value of 0.075 keV per channel as a requirement of applying the Multi-Group Analysis computer code (MGA) \cite{MGA}. The plutonium isotopic composition was determined by the commercial implementation of MGA, MGA++ \cite{MGA++}. The age is estimated by a software on the basis of the daughter/parent activity ratio of $^{241}$Am/$^{241}$Pu. The isotopic composition and the age of 8 Pu-Be sources, which are mainly the sources used in a previous paper \cite{CBM-NIM}, are shown in Table \ref{Tab:parameters}. The values obtained by MGA for the isotopic composition are in agreement with those obtained in Ref. \cite{CBM-NIM} by analyzing gamma- and x-rays below 300 keV. The age calculated by MGA agrees with the one which can be obtained from the date of production available in the records. 

\begin{figure}[htbp]
\begin{center}
\epsfig{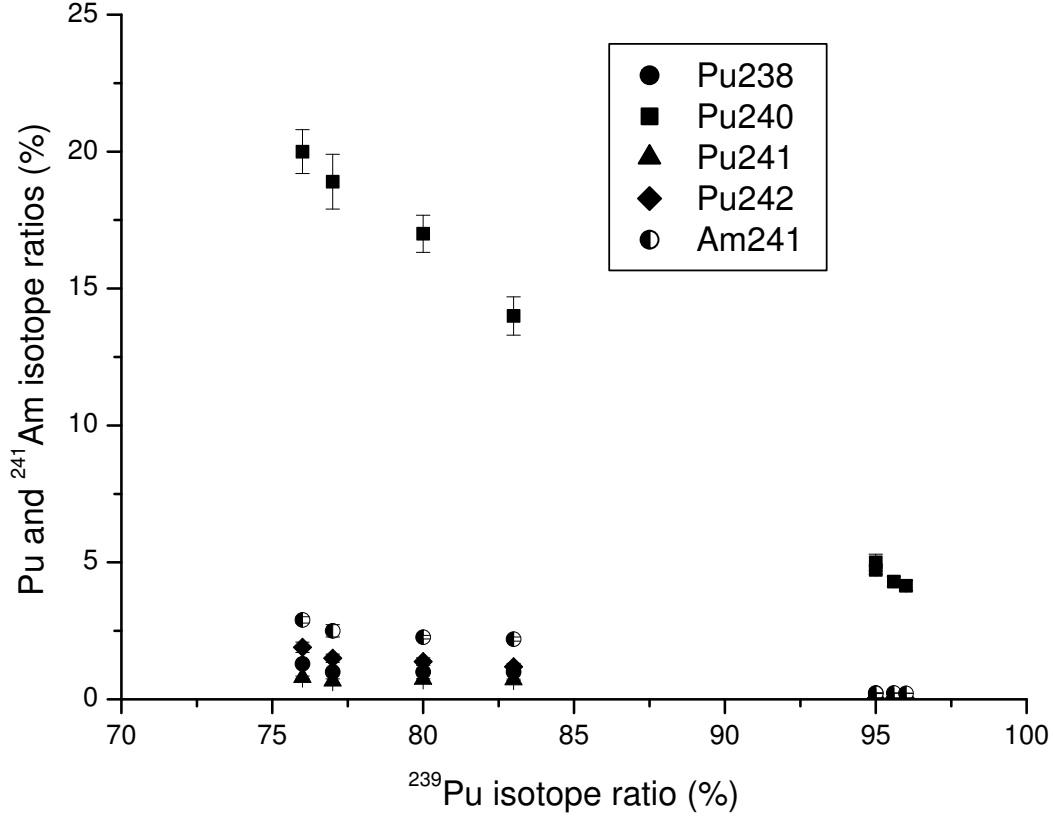}
\caption{Correlation between the $^{239}$Pu isotope ratio and isotope ratio of other Pu isotopes and $^{241}$Am}
\label{pu-correlations}
\end{center}
\end{figure}

The isotopic fractions of $^{238}$Pu, $^{240}$Pu, $^{241}$Pu, $^{242}$Pu and $^{241}$Am are plotted versus $^{239}$Pu isotopie ratio on Fig. \ref{pu-correlations}. It can be seen that there is a correlation between them. This correlation can be explained by the production of Pu isotopes in a nuclear power reactor.

\section{Determination of the Pu content}

\subsection{Calculating procedure}

\begin{figure}[htbp]
\begin{center}
\epsfig{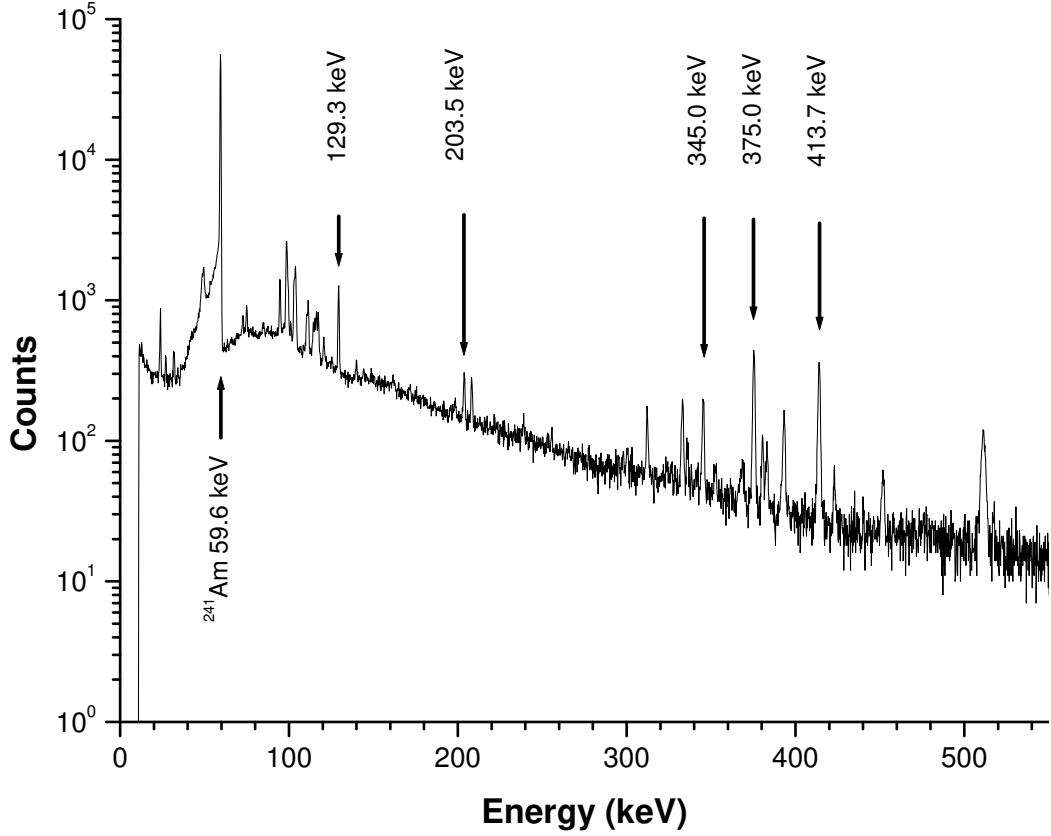}
\caption{Gamma spectrum of a Pu-Be source taken by a planar high-purity germanium detector}
\label{spectrum}
\end{center}
\end{figure}

A typical gamma spectrum of a Pu-Be source is shown in Fig. \ref{spectrum}, where the 129, 203, 345 and 413 keV lines of $^{239}$Pu can be clearly seen.

The correlation of the count rate of a gamma-ray with energy $E$, $C_E$, and total Pu content, $M_{Pu}$, can be expressed as

\begin{equation}
C_E(M_{Pu})=M_{Pu}f_{239}G_EO_E{1\over{F_E}}\ ,
\label{CE}
\end{equation}

where $f_{239}$  is the $^{239}$Pu isotopic ratio, $G_E$ is the specific gamma yield (gamma/gs) \cite{PANDA}, $O_E$ is the efficiency of detector measured by standard sources and $F_E$ is a correction factor for absorption. $F_E$ is the product of a factor accounting for self-absorption, $F_{1E}$, and a factor accounting for absorption in the steel wall of the container, $F_{2E}$. Assuming that the gamma-rays come parallel to the detector axis, $F_{1E}$ can be calculated using the far-field approximation for a cylindrical sample viewed along a diameter \cite{PANDA,self-shielding}:

\begin{equation}
F_{1E}={1 \over 2}{{\mu_{1E}D} \over{I_1(\mu_{1E}D)-L_1(\mu_{1E}D)}}\ , \label{F1}
\end{equation}
while

\begin{equation}
F_{2E}=\exp({\mu_{1CE}d_c})\ , \label{F2}
\end{equation}

where 
\begin{itemize}
\item $\mu_{lE}$ is the linear attenuation coefficient of the Pu-Be material, 
\item $D$ is the diameter of the sample, 
\item $I_1$ is modified Bessel function of order 1 \cite{functions},
\item $L_1$ is modified Struve function of order 1 \cite{functions}, 
\item  $\mu_{lCE}$ is the linear attenuation coefficient of the steel container taken from Ref. \cite{Veigele} and 
\item $d_C$ is the thickness of the container wall.
\end{itemize}

Furthermore,  $\mu_{lE}$ can be calculated from the mass attenuation coefficient (cm$^2$/g) of Pu,  $\mu_{PuE}$, and Be, $\mu_{BeE}$, \cite{Veigele} as:

\begin{equation}
\mu_{lE}=\rho\mu_E=\biggl( {239\over{239+9n}}\mu_{PuE}+ {{9n}\over{239+9n}}\mu_{BeE}\biggr)\ ,
\end{equation}

where $\rho$ is the density of the sample, $n$ is the ratio of the number of Be atoms to the number of Pu atoms (PuBe$_n$), and the numbers 239 and 9 are the atomic mass numbers of Pu and Be, respectively. In general, the Pu-Be material probably does not completely fill the inner volume of container. Nevertheless, it is expected to take on the cylindrical shape of the container, so the diameter $D$ of the Pu-Be material can be expressed as:

\begin{equation}
D(\rho,R,n,M_{Pu})=\biggl({{4M_{Pu}}\over{\pi\rho R}}\times{{239+9n}\over{239}}\biggr)^{1/3}\ ,
\end{equation}

where $R=H/D$ is the ratio of height and diameter of the Pu-Be material (see Fig. \ref{setup}).

\begin{figure}[htbp]
\begin{center}
\epsfig{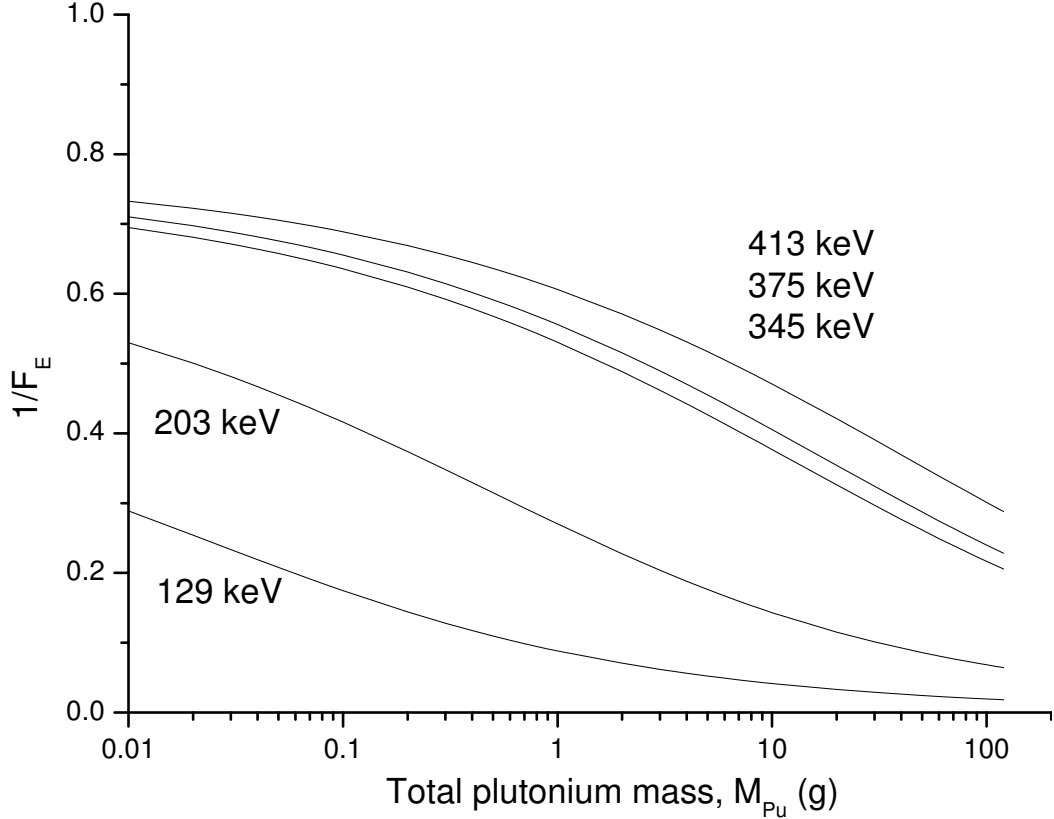}
\caption{Calculated values of $1/F_E$ versus total plutonium mass, $M_{Pu}$, for five $^{239}$Pu gamma-rays.}
\label{Fig:1perF}
\end{center}
\end{figure}

\begin{figure}[htbp]
\begin{center}
\epsfig{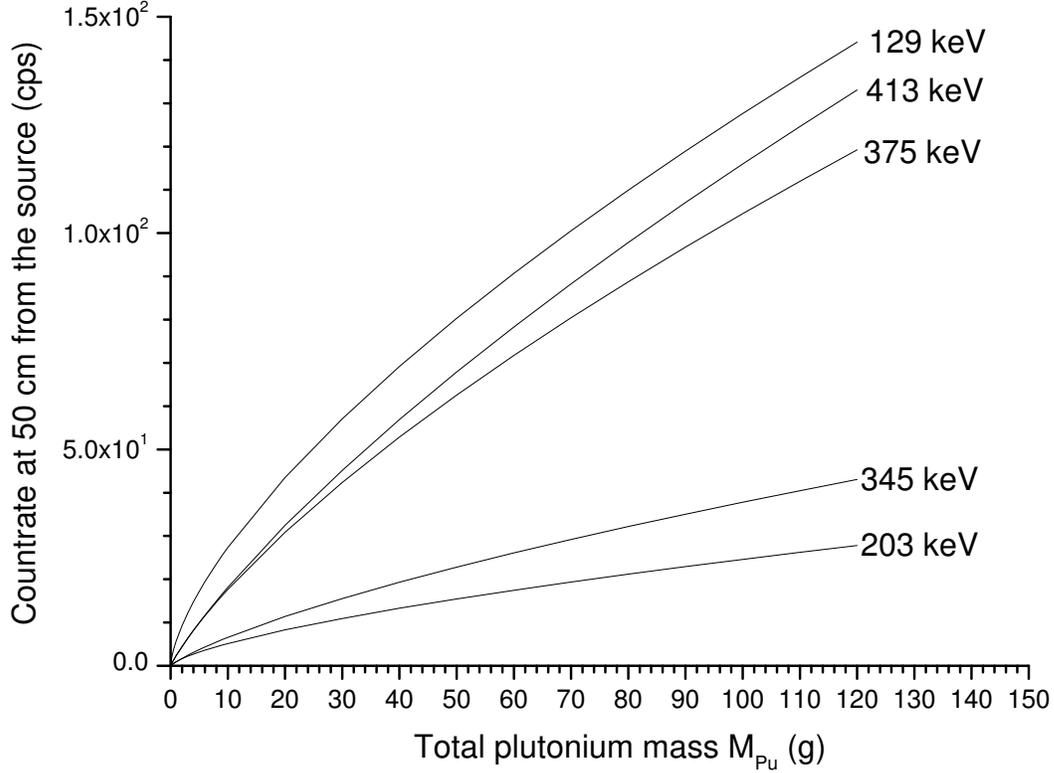}
\caption{Calculated count rate of the peaks of $^{239}$Pu at 50 cm distance from the source.}
\label{Fig:countrate}
\end{center}
\end{figure}

The calculated values of $F_E=F_{1E}\times F_{2E}$, $1/F_E$ and $C_E$ for the 129, 203, 345, 375 and 413 keV lines are plotted versus plutonium mass on figures \ref{Fig:FE}, \ref{Fig:1perF} and \ref{Fig:countrate}, respectively, in the range below 120 g. The values of the parameters were set as $n=13$, $\rho=3.7$ g/cm$^3$ for a ``usual'' Pu-Be material, and it was assumed that $R=1$, $d_C=4$ mm, and $f_{239}=1$ for a ``model'' source. Furthermore, $L=50$ cm was taken as a typical measurement distance, while $I_1$ and $L_1$ were taken from Ref. \cite{functions}. Using these values for the parameters, the following observations can be derived from formula (\ref{CE}) and Figs. \ref{Fig:FE}-\ref{Fig:countrate}:

\begin{figure}[htbp]
\begin{center}
\epsfig{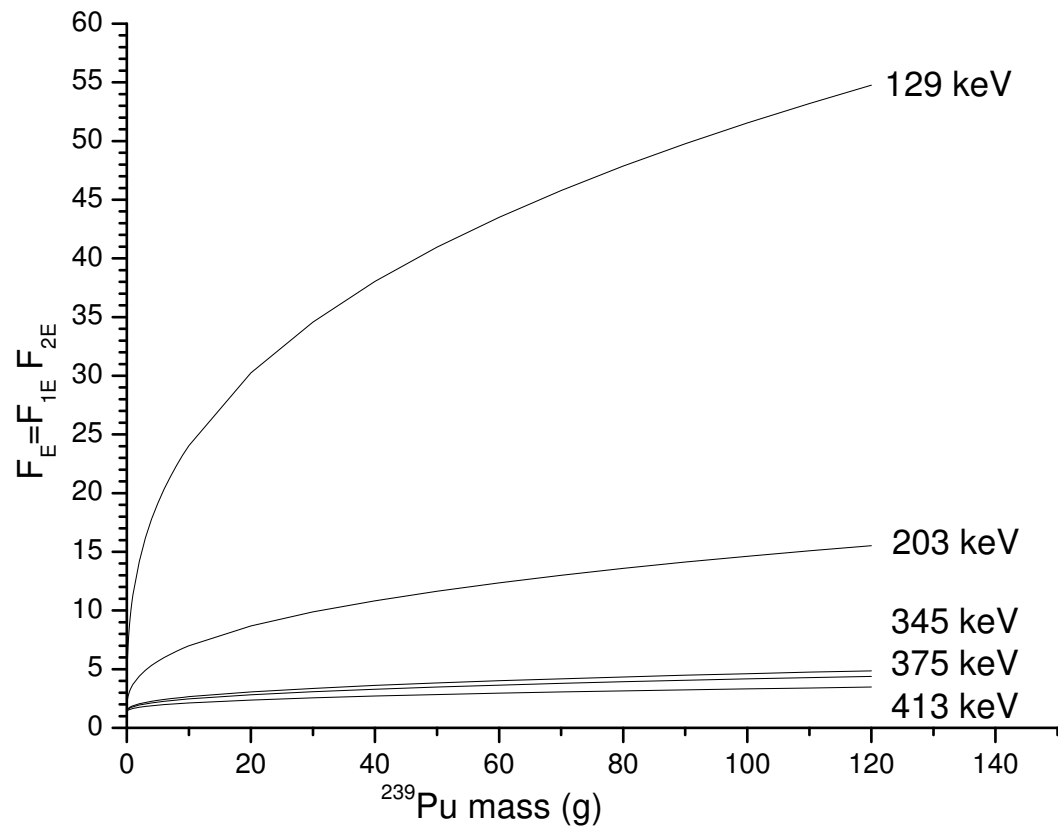}
\caption{Dependence of the correction factor for absorption, $F_E$, on the mass of $^{239}$Pu, assuming a cylindrical sample.}
\label{Fig:FE}
\end{center}
\end{figure}

The value of $F$ is relatively small for the 345, 375 and 413 keV lines and it varies slowly between 1.5 and 3 for plutonium masses in the range 0-120 g. Consequently, about 30-70 \% of the gama photons at these energies leaves the source.
On the other hand, for the 129 keV and 203 keV lines, the value of $F$ is relatively large. Furthermore, it rises very quickly (from 3 to 80 for 129 keV and from 2 to 10 for 203 keV for plutonium masses in the range 0-120 g) and only a small percentage of these gamma rays escapes from the source (1-30 \% for 129 keV and 10-50 \% for 203 keV). Note finally, that the count rate of the 129, 375 and 413 keV lines is much higher than that of the 203 and 345 keV lines. The first group will be called ``intensive group''.

For determining the plutonium content the use of gamma peaks of relatively high intensity is preferred, the count rate of which can be measured with a satisfactory statistical error within a reasonable measurement time. Namely, since neutron damage to the detector crystal is to be reduced to the minimum, as short a measurement time as possible should be chosen. Furthermore, the count rate of the chosen gamma peaks should not be very sensitive to the geometrical dimensions of the assayed sample, because the latter are only known with a relatively large uncertainty. The gamma peaks at
375 and 413 keV satisfy both these requirements. The count rate of the peak at 129 keV is also high, but it strongly depends on the geometrical dimensions of the sample, so in the present work it has not been used for the quantitative assay of the samples. Note, however, that exactly this property of the 129 keV line can be used for verifying the parameters of the source, i.e., for estimating the ratio $R=H/D$. This aspect will be reported in a subsequent paper.

Equation (\ref{CE}) can be rewritten in the form
\begin{equation}
M_{Pu}={{C_E}\over{f_{239}G_E O_E}}F_E=f(C_E,f_{239},G_E,O_E,P)\ ,
\label{Pumass}
\end{equation}
where $P$ is the set \{$n$, $\rho$, $R$, $d_C$\} of parameters of the source. Using formula (\ref{Pumass}) $M_{Pu}$ will be derived from the count rates $C_E$ by taking an adopted value for the parameters in the set $P$. It means that for each source there will be more values for $M_{Pu}$, calculated from the count rate of gamma peaks at different energies. In the present case we will only consider the values of $M_{Pu}$ which can be obtained from the count rates of the gamma peaks at 375 and 413 keV. The two values should agree with each other and the final value of $M_{Pu}$ will be calculated as the weighted average of these two, for each source separately.

\subsection{Analyzing the sources of errors}

Although, in general, each Pu-Be neutron source has a certification document providing the parameters of the container (see Table \ref{Tab:parameters}), the values in the parameter set $P$ are not always exactly known. The error of $M_{Pu}$, $\Delta M$, caused by the uncertainties of these parameters can be estimated as
\begin{equation}
\Delta M={{\Delta C_E}\over{f_{239}G_E O_E}}F_E=
{{F_E}\over{f_{239}G_E O_E}}\sqrt{\sum_k \biggl({{\partial C_E}\over{\partial P_k}}\biggr)^2(\Delta P_k)^2}\ ,
\label{error}
\end{equation}
where $P_k$ denotes the $k$-th element in the parameter set $P$.
The uncertainties associated with systematical errors of $G_E$, $O_E$ and $F_E$ are ignored in the above equation, because in this section we wish to concentrate only on the sources of error connected to the inappropriate knowledge of the parameter set $P$.

\begin{figure}[htbp]
\begin{center}
\epsfig{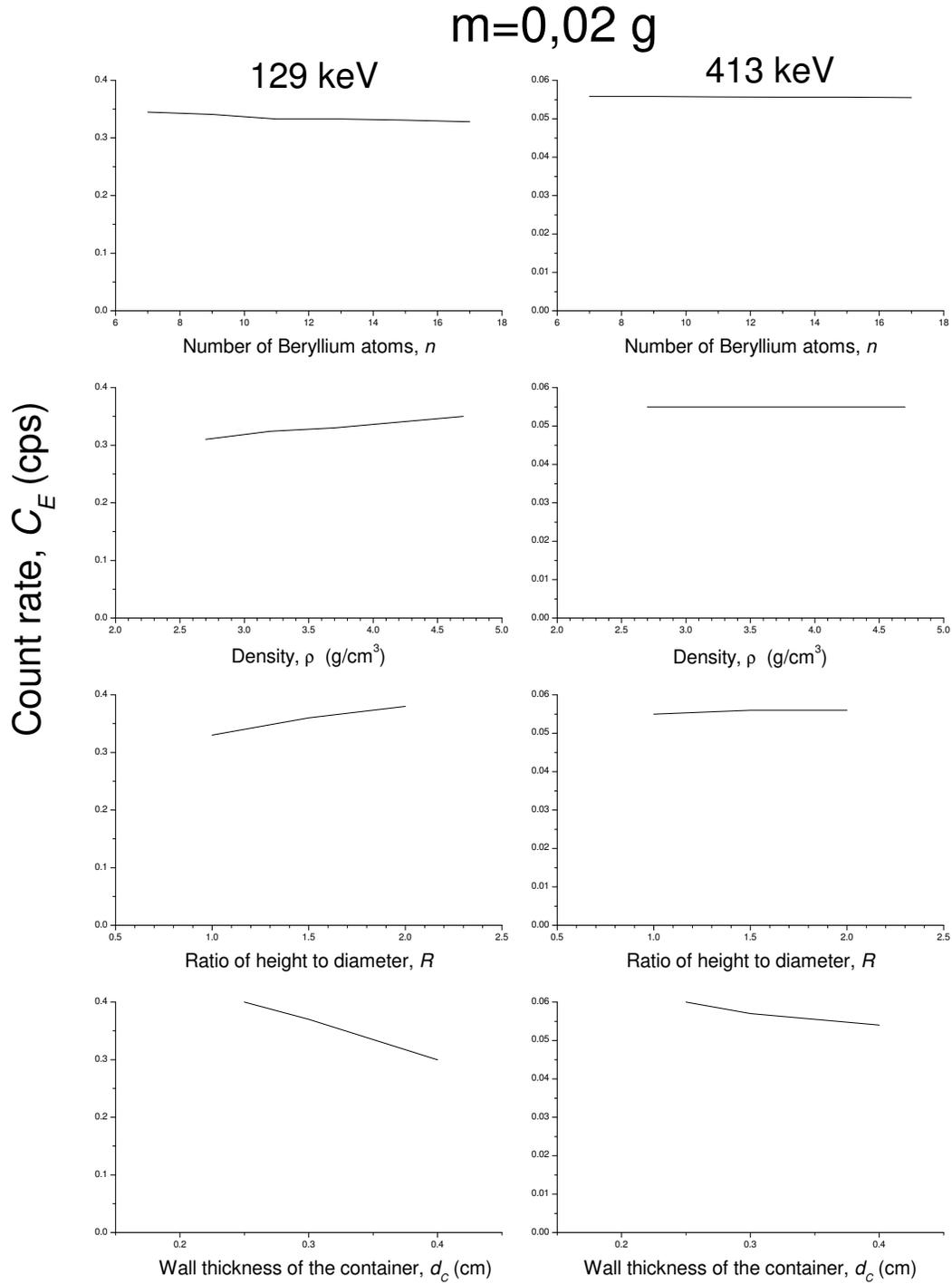}
\caption{Calculated count rate of the 129 and 413 keV peaks of $^{239}$Pu at 50 cm distance from the source as a function of various parameters of the source with total plutonium mass $M_{Pu}=0.02$ g.}
\label{Fig:CE002g}
\end{center}
\end{figure}

\begin{figure}[htbp]
\begin{center}
\epsfig{figure=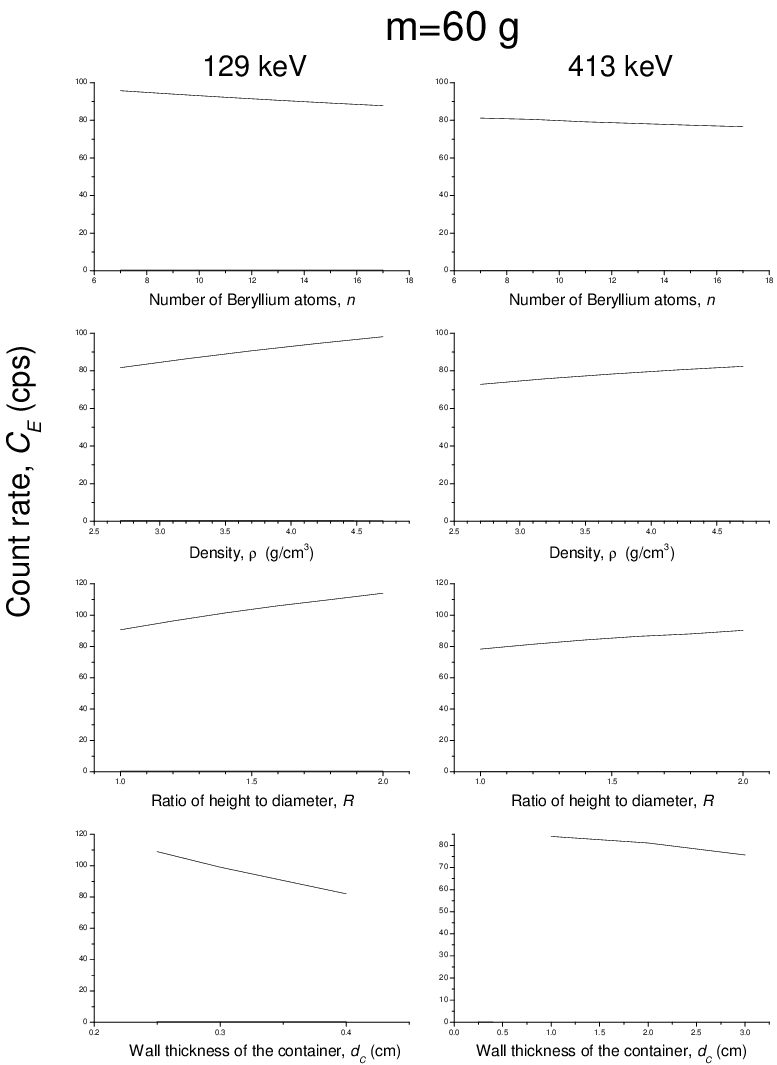, width=\textwidth}
\caption{Calculated count rate of the 129 and 413 keV peaks of $^{239}$Pu at 50 cm distance from the source as a function of various parameters of the source with total plutonium mass $M_{Pu}=60$ g.}
\label{Fig:CE60g}
\end{center}
\end{figure}

The values of $C_E$ calculated according to equation (\ref{CE}) are plotted on Figs. \ref{Fig:CE002g} and \ref{Fig:CE60g} for the 
129 and 413 keV lines as functions of the parameters $n$, $\rho$, $R$ and $d_C$, for two typical $M_{Pu}$ values of 0.02 g and 60 g. The values of $C_E$ for 413 kev vary very slowly with $n$ and $\rho$, and a bit more quickly with $R$ and $d_C$. This means that the values of $\partial C_E/\partial P_k$ in equation (\ref{error}) are small in the case of 413 keV. The same is true for 375 keV as well.  Therefore, the total Pu content can be estimated from the count rate of the 375 and 413 keV peaks with high accuracy even if one uses the uncertain values of the parameters given in the certificate of the Pu-Be source. A rough estimation based on the calculation model presented above and illustrated in Figs. \ref{Fig:CE002g} and \ref{Fig:CE60g}, using typical parameters of the sources, yields that the relative error of $M_{Pu}$ caused by the uncertainties of these parameters is below 5 \%. 
 
The value of $C_E$ for the 129 keV peak is more sensitive to the changes of $R$ and $d_C$, thus the uncertainties of $R$ and $d_C$ cause a larger error in the value of $M_{Pu}$ calculated using the count rate of the 129 keV peak. Therefore, as already remarked, in this work the 129 keV peak will not be used for estimating the plutonium content.

\subsection{Determination of the Pu content}

The parameters of the sources, the measurement set up, the count rates and the Pu contents calculated using formula (\ref{Pumass}) are presented in Tables \ref{Tab:parameters}-\ref{Tab:Pucontent}. The final Pu contents (Table \ref{Tab:Pucontent}, column 6) are the averages of the Pu contents estimated from the count rates of the 375 and 413 keV lines.

\section{Results and discussion}

The Pu contents obtained by the described method are smaller by about 30-50\% than the values obtaied previously for the same Pu-Be sources by the method which uses the isotopic composition and the total neutron output (the ``combined method'' or CBM \cite{CBM-NIM}) for calculating the Pu content. This implies that the values of the specific neutron yield are larger than the ones used in the CBM, by about 30-50\%. If, in addition, the number of neutrons from ($\alpha$,n) reactions is estimated, for example, by measuring the 4438 keV gamma line \cite{Drake}, then the specific neutron yield can also be determined, using the results of the present work. 

The calibration curve for the $M-R/T$ correlation of the ``R/T'' method mentioned in the introduction of this paper becomes more accurate if one uses the total Pu contents obtained by the present method instead of the ones obtained by the CBM.

In each measurement, the detector was exposed to an integral neutron flux of about $10^4-10^5$ n/cm$^2$ to get a statistical error below 5\% for the peaks in the ``intensive group'', IG. The measurement time should be much longer if the statistical error of all five relevant gamma-rays of $^{239}$Pu is to be bellow 5\%. In the described measuremts the degree of neutron damage to the detector is expected to be negligible, because in the present measurements the integral neutron flux was much smaller than $10^9$ n/cm$^2$, which is the integral flux that causes a noticeable damage to the detector, as estimated by some authors \cite{Kraner,Zachary}.

The plutonium content of the sources can also be estimated by the the so-called ``infinite energy method'' \cite{infinite} in the following way. The value of $m(1/E)=C_E/(f_{239}G_EO_E)$ is plotted as a function of $1/E$ for five gamma-rays (129, 203, 345, 375 and 413 keV). The curve fitted to these points is then extrapolated to $1/E=0$, and the intercept, $m(1/E=0)$ yields the plutonium content of the source $M_{Pu}$. Comparing $m(1/E)$ to formula (\ref{CE}), it can be seen that $m(1/E)=M_{Pu}/F_E$. Therefore the relation between the method presented in this paper and the infinite energy method is that in this work $F_E$ was used to correct for self-absorption, instead of using the extrapolation of the infinite energy method. Calculating $F_E$ requires the knowledge of certain parameters of the sources, such as, e.g., the dimensions of the source, the type of the container etc., whereas the infinite energy method does not. However, if the required data are available, the method presented here gives a more accurate result for the plutonium content than the infinite energy method. In the present case these data were available from the certificates of the assayed sources, so their plutonium content was evaluated using the method described in this work. An exception is the source having the identifier 4.7-555 (Table \ref{Tab:Pucontent}) which was encased in a double container of unknown dimensions, so its plutonium content was estimated using the infinite energy method.

Alternatively, if the source dimensions are not exactly known, the 129 keV line can be used for monitoring the parameters of the sources. Presently a method based on using the 129 keV line is being developed in order to determine the plutonium content of completely unknown Pu-Be sources with high precision.

\section*{Note added in the revised version}

After the original version of this paper has been submitted, the results of the gamma-spectrometric plutonium-mass measurements of several Pu-Be sources were confirmed by calorimetric measurements \cite{ISPRA-ESARDA}. The comparison of the two methods will be dealt with in a subsequent paper.

\section*{Acknowledgement}

The research on the verification of Pu-Be sources was accepted by the IAEA as part of the Hungarian support programme to the IAEA. This work was partially supported by the Hungarian Atomic Energy Authority under contract No. OAH-ABA-ÁNI-04/03. The author wishes to thank Jozsef Zsigrai for his help in revising the manuscript.

\begin{landscape}

\section*{Tables}

\begin{table}[htbp]
\caption{Parameters of the assayed Pu-Be sources taken from the certification documents and the parameters of measurements. The first number in the source identifier is its nominal Pu content in grams. (*Put in an aluminum container.)}
\begin{tabular}{lllllllll}
\hline
Identifyer & \parbox{2cm}{Neutron output (n/s)} & \multicolumn{2}{c}{\parbox{3.2cm}{Outer dimensions of the container (cm)}} & \multicolumn{2}{c}{\parbox{3.2cm}{Inner dimensions of the container (cm)}} & \multicolumn{3}{c}{Measurement parameters} \\ 
\cline{3-9}
   &  & Diameter & Height & Diameter, D & Height, H & \parbox{1cm}{L (cm)} & \parbox{2cm}{Meas. time (s)} & \parbox{2.5cm}{Neutron flux ($\times 10^4$ n/cm$^2$)} \\ 
\hline
  0.2-480 & $1.1\times 10 ^4$ & $1.0\pm 0.02$ & $1.9\pm 0.05$ &  &  & 20 & 680.0 & 0.15 \\ 
  2-479 & $1.1\times 10 ^5$ & $1.5\pm 0.02$ & $2.4\pm 0.02$ &  &  & 50 & 1352.85 & 0.47 \\ 
  1.9-611a & $1.16\times 10 ^5$ & 2.0 & 3.0 & 1.3 & 1.95 & 50 & 1198.9 & 0.44 \\ 
  4.7-555* & $2.9\times 10 ^5$ & 2.0 & 3.0 & 1.3 & 1.95 & 50 & 695.5 & 0.59 \\ 
  4-407 & $2.68\times 10 ^5$ & 2.0 & 3.0 & 1.3 & 1.95 & 50 & 974.0 & 1.02 \\ 
  37-425 & $2.26\times 10 ^6$ & $2.4\pm 0.02$ & 3.0 & 1.6 & 1.60 & 200 & 3326 & 5.71 \\ 
  178-461 & $1.1\times 10 ^7$ & 3.5 & 4.5 & 2.7 & 2.70 & 200 & 606.5 & 1.33 \\ 
  85-701 & $5.27\times 10 ^6$ & 3.5 & 4.2 & 2.9 & 3.15 & 200 & 648.8 & 0.68 \\ 
\hline
\end{tabular}
\label{Tab:parameters}
\end{table}

\begin{table}[htbp]
\caption{Isotopic composition and age of the sources determined by the MGA code from the gamma spectra taken in the region 0-600 keV.}
\begin{tabular}{llllllllll}
\hline
Identifyer & \multicolumn{6}{c}{Isotope ratio (\%)} & Age (years) & \parbox{1.7cm}{Date on\\ certificate} & \parbox{1.9cm}{Measure\-ment date} \\ 
\cline{2-7}
 & $^{238}$Pu & $^{239}$Pu & $^{240}$Pu & $^{241}$Pu & $^{242}$Pu & $^{241}$Am &  &  &  \\ 
\hline
0.2-480 & 1$\pm$3.5\% & 83$\pm$1\% & 14$\pm$5\% & 0.71$\pm$4.4\% & 1.19$\pm$10\% & 2.2$\pm$3.5\% & 28.4$\pm$0.6 & 1982.10 & 2002.07.25 \\ 
2.479 & 1$\pm$3\% & 79.9$\pm$1\% & 17$\pm$4\% & 0.73$\pm$3.5\% & 1.38$\pm$10\% & 2.27$\pm$3\% & 29.7$\pm$0.5 & 1997.10 & 2002.07.25 \\ 
1.9-611a & 0.004$\pm$50\% & 95.6$\pm$0.2\% & 4.3$\pm$4\% & 0.046$\pm$5.5\% & 0.014$\pm$10\% & 0.23$\pm$2\% & 37.6$\pm$1 & 1965.06.08 & 2002.06.25 \\ 
4.7-555 & 0.0035$\pm$75\% & 96$\pm$0.2\% & 4.15$\pm$4.5\% & 0.05$\pm$6.5\% & 0.012$\pm$10\% & 0.22$\pm$2.5\% & 35.5$\pm$1.3 & 1967.03.16 & 2002.02.23 \\ 
4-407 & 0.001$\pm$200\% & 95$\pm$0.2\% & 4.73$\pm$3.5\% & 0.04$\pm$6.5\% & 0.015$\pm$10\% & 0.23$\pm$2\% & 39$\pm$1.3 & 1965.01.30 & 2002.11.19 \\ 
37-425 & 1.06$\pm$3\% & 77$\pm$1\% & 19.7$\pm$4\% & 0.7$\pm$4\% & 1.6$\pm$10\% & 2.8$\pm$3\% & 33.5$\pm$0.5 & 1973.10.29 & 2004.12.09 \\ 
178-461 & 1.3$\pm$4\% & 76$\pm$1.5\% & 20$\pm$4\% & 0.8$\pm$6\% & 1.9$\pm$10\% & 2.9$\pm$4\% & 31.5$\pm$0.5 & 1976.08.26 & 2002.04.28 \\ 
85-701 & 0.0076$\pm$56\% & 95$\pm$0.3\% & 5$\pm$6\% & 0.044$\pm$12\% & 0.017$\pm$10\% & 0.22$\pm$3.5\% & 37$\pm$1.5 & 1966.08.20 & 2002.07.25 \\ 
\hline
\end{tabular}\label{Tab:isotopic}
\end{table}

\end{landscape}

\begin{table}[htbp]
\caption{Parameters used for calculations}
\begin{tabular}{llllll}
\hline
Gamma-energy & 129 keV & 203 keV & 345 keV & 375 keV & 413 keV \\ 
\hline
$\mu_E$ (cm$^2$/g) \cite{Veigele}& 2.05 & 0.9 & 0.3 & 0.27 & 0.205 \\ 
$\mu_l({\rm Fe})$ (cm$^{-1}$) \cite{Veigele}& 1.85 & 1.15 & 0.81 & 0.77 & 0.73 \\ 
$G_E$ ($10^4$ photons/gs) \cite{PANDA}& 14.361 & 1.2847 & 1.2829 & 3.6014 & 3.6 \\ 
$O_E\times 10^4$ (measured at 50 cm) & 4.6 & 2.8 & 1.36 & 1.21 & 1.07 \\ 
\hline
\end{tabular}\label{Tab:parameters2}
\end{table}

\bigskip

\begin{table}[htbp]
\caption{Count rate of \textsuperscript{239}Pu gamma-rays}
\begin{tabular}{llllll}
\hline
Identifier & \multicolumn{5}{c}{Count rate (cps)}\\
\cline{2-6}
& 129 keV&203 keV&345 keV&375 keV&413 keV\\
\hline
0.2{}-480
&2.25$\pm$0.12&0.191$\pm$0.059&0.159$\pm$0.041&0.31$\pm$0.059&0.307$\pm$0.059\\
2-479&1.83$\pm$0.11&0.239$\pm$0.037&0.174$\pm$0.032&0.448$\pm$0.04&0.427$\pm$0.044\\
1.9-611a&5.88$\pm$0.13&1.053$\pm$0.083&0.799$\pm$0.05&2.08$\pm$0.06&1.92$\pm$0.063\\
4.7-555&13.63$\pm$0.36&2.21$\pm$0.21&2.21$\pm$0.15&5.19$\pm$0.18&4.66$\pm$0.15\\
4-407&14.55$\pm$0.21&2.36$\pm$0.12&2.15$\pm$0.08&5.75$\pm$0.13&5.21$\pm$0.12\\
37-425&0.85$\pm$0.05&0.15$\pm$0.03&0.14$\pm$0.015&0.38$\pm$0.02&0.375$\pm$0.017\\
178-461&2.67$\pm$0.26&0.53$\pm$0.16&0.63$\pm$0.13&1.53$\pm$0.15&1.52$\pm$0.10\\
85-701&5.79$\pm$0.29&1.15$\pm$0.15&1.35.$\pm$0.12&3.96$\pm$0.15&4.02$\pm$0.15\\
\hline
\end{tabular}
\label{Tab:countrate}
\end{table}

\bigskip

\begin{table}[htbp]
\caption{Total plutonium content estimated from the count rate of the 375 and 413 keV lines.}
\begin{tabular}{llllll}
\hline
Identifier&\multicolumn{2}{c}{Parameters}&\multicolumn{3}{c}{Total plutonium content (g)}\\
\cline{2-6}
&R&d$_C$ (cm)&375 keV&413 keV&Average\\
\hline
0.2-480&
1.9&0.35&0.019$\pm$0.004 &0.021$\pm$0.004&0.020$\pm$0.004\\
2-479&1.6&0.35&0.20$\pm$0.02&0.20$\pm$0.02&0.205$\pm$0.04\\
1.9-611a&1.5&0.35&0.85$\pm$0.03&0.83$\pm$0.03&0.84$\pm$0.03\\
4.7-555&\multicolumn{2}{c}{unknown}&\multicolumn{2}{c}{infinite energy method}&3$\pm$0.4\\
4-407&1.5&0.35&2.65$\pm$0.07&2.47$\pm$0.06&2.55$\pm$.05\\
37-425&1&0.4&4$\pm$0.25&4$\pm$0.2&4.0$\pm$0.2\\
178-461&1&0.4&19.3$\pm$2.3&19.5$\pm$2.2&19.5$\pm$1.6\\
85-701&1.08&0.3&49.5$\pm$2.5&49.1$\pm$2.3&49.2$\pm$1.5\\
\hline
\end{tabular}
\label{Tab:Pucontent}
\end{table}

\end{document}